\documentclass[aps,prd,onecolumn,superscriptaddress,showpacs,floatfix,10pt]{revtex4}
\usepackage{graphicx}
\usepackage{amssymb,amsmath}
\usepackage{subfigure}
\usepackage{epsfig}

\begin{document}

\title{Quantum dynamics of particles in a discrete two-branes world model: Can matter particles exchange occur between branes?}


\author{Micha\"{e}l Sarrazin}
\email{michael.sarrazin@fundp.ac.be} \affiliation{Laboratoire de
Physique du Solide, Facult\'es Universitaires Notre-Dame de la
Paix, \\61 rue de Bruxelles, B-5000 Namur, Belgium}

\author{Fabrice Petit}
\email{f.petit@bcrc.be} \affiliation{Belgian Ceramic Research
Centre,\\4 avenue du gouverneur Cornez, B-7000 Mons, Belgium}

\begin{abstract}
In a recent paper, a model for describing the quantum dynamics of
massive particles in a non-commutative two-sheeted spacetime was
proposed. This model considers a universe made with two spacetime
sheets embedded in a 5D bulk where the fifth dimension is
restricted to only two points. It was shown that this construction
has several important consequences for the quantum dynamics of
massive particles. Most notably, it was demonstrated that a
coupling arises between the two sheets allowing matter exchange in
presence of intense magnetic vector potentials. In this paper, we
show that non-commutative geometry is not absolutely necessary to
obtain such a result since a more traditional approach allows one
to reach a similar conclusion. The fact that two different
approaches provide similar results suggests that standard matter
exchange between branes might finally occur contrary to
conventional belief.
\end{abstract}


\maketitle

\section{Introduction}

The idea that our four dimensional spacetime is only a part of an
extended multidimensional universe is a recurrent topic in
literature. The idea traces back to the 1920's, to the works of
Kaluza [1] and Klein [2] who tried to unify electromagnetism and
gravitation by assuming that the photon originates from the fifth
component of the metric. However, despite the interest of
unification, the physical predictions of the model revealed
unsustainable and the overall approach was abandoned for a while.
In recent few years however, there has been a renewed interest for
such multidimensional scenario in different contexts and using
different mathematics (see references [3] to [13] for instance).
An important breakthrough was undoubtedly made possible thanks to
the emergence of non-commutative geometry (NCG) [14]. For the sake
of simplicity, NCG is generally used to describe spacetimes
composed of a continuous part (a four dimensional hypersurface)
times a discrete part where non-commutativity acts
[13,14,15,16,17]. Several extensions of general relativity have
thus been proposed over the years. It has been shown that those
models which can be seen as minimal extensions of present theories
could give nice explanations to several puzzling phenomena, the
most important one being the so-called hierarchy problem. In most
approaches, spacetime is assumed to be two-sheeted. The two sheets
where left and right fermions live are embedded in a discrete five
dimensional space simply reduced to two points. The superiority of
non-commutativity arises precisely by the way it elegantly allows
to make mathematics in the discrete space. The idea of a doubled
spacetime is also in agreement with some extensions of the
standard model postulating the existence of a mirror sector (or
hidden sector, depending of the approach) to explain parity
violation problems [15,16]. Yet, in spite of its powerfulness, the
use of non-commutativity is still very limited in physics. The
reason arises from the mathematical formalism which makes
construction of new theories a hard task. A possible alternative
is to keep the idea of discrete dimensional space while
eliminating non-commutativity. Several papers have demonstrated
that it is indeed possible to get similar results to NC approach
without recourse to its formalism [9,10,11,12]. For instance,
Kokado and coworkers have shown that it is possible to derive pure
Einstein action on $M4 \times Z2$ geometry (leading to Brans-Dicke
theory in four dimensional spacetime) simply by redefining the
notions of parallel transport and Riemann curvature tensors in a
way appropriate to those spacetimes [9]. More recently,
Arkani-Hamed and al. [10] have developed the idea of discrete
gravitational dimensions and developed an effective field theory
for massive gravitons. Those spaces are defined by four
dimensional sites in a compact discretized space taking the form
of either a circle or an interval (in five dimensions). Each site
is then endowed with its own four dimensional metric and the 5D
Einstein-Hilbert action is simply discretized using a finite
difference method. It is also the path followed by Defayet and al.
in their approach to multigravity [11,12]. Those authors have
shown that different discretization schemes are even possible,
some of them leading to theories free of ghosts and usual
complications due to the recourse of a lattice extra space (at
least at the linear level of approximation). So, it appears that a
simple discretization of extra dimensions although simple could be
useful to study.

In this paper we are following a similar approach applied to the
fundamental equations of the quantum domain instead of general
relativity. Hence, we are considering a five dimensional spacetime
where the fifth dimension, a segment, is restricted to only two
points. Relevant extensions of the classical Dirac's et Pauli's
equations are then derived and the effects of the existence of a
second spacetime sheet on the dynamics of massive particles are
studied. In a previous paper [18], a similar work was proposed but
a subsequent use of NCG was necessary to build the model. Besides,
the approach we follow here is mathematically simplified, it will
be shown that, except for secondary aspects, the basic results of
the model are similar to those obtained with a non-commutative
(NC) formalism. The results of the present paper appear also to be
somewhat clearer. Nevertheless, it is shown that both approaches
predict an electromagnetic coupling between the two sheets that
might have dramatic experimental consequences. Indeed, since the
particles wave functions are now five dimensional, particles are
simultaneously present on both sheets although with different
probabilities of presence. It is demonstrated that a high
electromagnetic vector potential can modify those probabilities
such that a particle initially localized in the first sheet can be
transferred into the second sheet. The differences between NC and
classical approach are also reviewed. The most noticeable ones
concerns the confinement of the particles in their spacetime
sheets. NC approach predicts that without electromagnetic vector
potential, the particles are perfectly stable and remain in their
original sheets. Contrarily, the "finite difference" approach
suggests that particles oscillate between the two sheets with a
time periodicity depending on the distance between the sheets.
Hence, the observation of particle disappearance in laboratory
conditions could provide a way of determining whether the universe
is doubled and non-commutative or not.

\section{The Model}

Let us consider a quantum model for a two-branes universe based on
a non-trivial generalization of the Dirac equation. In this model,
the fifth dimension is reduced to two points with coordinates $\pm
\delta /2$. The branes are assumed to be located at those points
and $\delta $ is the distance between the two branes. From our
point of view, this distance should be considered as a
phenomenological one. More precisely, in the context of a
two-branes universe, the $M4 \times Z2$ manifold representation is
a convenient approach to formalize the two-branes world problem.
The $Z2$ dimension has not necessarily a physical existence and
can be considered just as an abstract dimension. As a consequence,
we must take care about the fact that the distance mentioned in
the present work does not directly correspond to the concept of
distance between branes often mentioned in previous works. Hence,
the fifth dimensional generalization of the covariant Dirac
equation can be written as

\begin{equation}
i\gamma ^\mu \partial _\mu \psi +i\gamma ^5\partial _5\psi -m\psi
=0 \label{3}
\end{equation}

where the matrix $\gamma ^5=i\gamma ^0\gamma ^1\gamma ^2\gamma ^3$
anticommutes with the usual Dirac gamma matrices $\gamma ^\mu $
such that $\mu =0,1,2,3 $. Now, let us define $\psi _{+}$
(respectively $\psi _{-}$) the wave function at the point $+\delta
/2$ (respectively $-\delta /2$). Then, the discrete derivative
$\partial _5\psi $ can be simply written as a finite difference
involving $\psi _{+}$ and $\psi _{-}$ through

\begin{equation}
(\partial _5\psi )_{\pm }=\pm g(\psi _{+}-\psi _{-}) \label{4}
\end{equation}

with $g=1/\delta $. Note that the NCG [13,14,15,16,17] also uses a
discrete derivative along the fifth dimension to generalize the
classical Dirac operator in the case of a two-sheeted spacetime.
Our approach is however quite different and in fact, no
non-commutative mathematics will be explicitly used throughout
this paper. Using the discrete derivative, it can be shown that
the 5D Dirac equation breaks down into a set of two coupled
differential equations

\begin{equation}
\left\{
\begin{array}{c}
i\gamma ^\mu \partial _\mu \psi _{+}+ig\gamma ^5\psi _{+}-ig\gamma
^5\psi
_{-}-m\psi _{+}=0 \\
i\gamma ^\mu \partial _\mu \psi _{-}+ig\gamma ^5\psi _{-}-ig\gamma
^5\psi _{+}-m\psi _{-}=0
\end{array}
\right.  \label{5}
\end{equation}
which can be rewritten in a more compact form using a matrix
formalism, i.e.
\begin{equation}
\left\{ i\Gamma ^\mu \partial _\mu +ig\Gamma ^5-m\right\} \Psi =0
\label{6}
\end{equation}
where
\begin{equation}
\Psi =\left(
\begin{array}{c}
\psi _{+} \\
\psi _{-}
\end{array}
\right) \label{7}
\end{equation}
with $\Gamma ^\mu =\left[
\begin{array}{cc}
\gamma ^\mu & 0 \\
0 & \gamma ^\mu
\end{array}
\right] $ and $\Gamma ^5=\left[
\begin{array}{cc}
\gamma ^5 & -\gamma ^5 \\
-\gamma ^5 & \gamma ^5
\end{array}
\right] .$

We get
\begin{equation}
\left[ \Gamma ^\mu ,\Gamma ^5\right] =0 \label{8}
\end{equation}
and
\begin{equation}
\Gamma ^{5^{\ 2}}=2\Gamma =2\left[
\begin{array}{cc}
{\bf 1}_{4\times 4} & -{\bf 1}_{4\times 4} \\
-{\bf 1}_{4\times 4} & {\bf 1}_{4\times 4}
\end{array}
\right] \label{9}
\end{equation}
Then, it is straightforward to show that the Lagrangian associated
with eq.4 takes the form

\begin{equation}
{\cal L}=\overline{\Psi }\left\{ i\Gamma ^\mu \partial _\mu
+ig\Gamma ^5-m\right\} \Psi \label{10}
\end{equation}
where

\begin{equation}
\overline{\Psi }=\left(
\begin{array}{cc}
\overline{\psi }_{+} & \overline{\psi }_{-}
\end{array}
\right) \label{11}
\end{equation}

We see that our model describes two interacting Dirac's fields,
each one being related to a specific brane. The interaction arises
from the $ig\Gamma ^5$ term which equals to zero in the case of
infinitely separated sheets. We stress that ${\cal L}$ is the
simplest non trivial Lagrangian relevant for describing quantum
interactions between different branes. It can be verified that
${\cal L}$ is $CPT$, $PT$ and $C$ invariant but it is not $P$ and
$T$ invariant. It would be interesting to study if the fixed time
arrow of our spacetime could be linked to this broken symmetry. If
this assumption holds, then it would be very tempting to assume an
opposite time arrow in the other sheet to restore symmetry. At
this stage, we just underline that if we suppress the diagonal
terms $\gamma^5$ from $\Gamma^5$, we then retrieve the equations
obtained for a two-sheeted spacetime using the NC formalism [18].
We are now going to study the consequences of this subtle
difference between both approaches.

\section{Free field eigenmodes}

Let us determine the eigenmodes of the two-sheeted Dirac equation
in the free field case. The solution can be easily derived by
introducing a potential $\Phi $ such that

\begin{equation}
\Psi =\left\{ i\Gamma ^\mu \partial _\mu +ig\Gamma ^5+m\right\}
\Phi  \label{12}
\end{equation}
Using eq.4 and eq.10, it can then be demonstrated that $\Phi $
satisfies the equation
\begin{equation}
\left\{ \square +2g^2\Gamma +m^2\right\} \Phi =0 \label{13}
\end{equation}
such that the solutions $\Psi $ of eq.4 can be deduced from the
solutions $\Phi $ of eq.11. Let us look for solutions of the form

\begin{equation}
\Phi =\Phi _0e^{-i\varepsilon p.x}=\Phi _0e^{-i\varepsilon
(E_pt-{\bf p.x)}} \label{14}
\end{equation}
where $\varepsilon =+1$ for positive energies and $\varepsilon
=-1$ for negative ones. By replacing the expression of $\Phi $ in
eq.11, one gets

\begin{equation}
\left[
\begin{array}{cc}
-E_p^2+p^2+m^2+2g^2 & -2g^2 \\
-2g^2 & -E_p^2+p^2+m^2+2g^2
\end{array}
\right] \Phi _0=0  \label{15}
\end{equation}
which yields two solutions

The first solution is obtained in the case $E_p=\sqrt{p^2+m^2}$
(with $\varepsilon =\pm 1$) for which we get
\begin{equation}
\Phi _0\rightarrow \left[
\begin{array}{c}
\phi _{\varepsilon ,\lambda } \\
\phi _{\varepsilon ,\lambda }
\end{array}
\right]  \label{16}
\end{equation}

The second solution corresponds to
$\widetilde{E}_p=\sqrt{p^2+m^2+4g^2}$ (with $\varepsilon =\pm 1$)
for which we get
\begin{equation}
\Phi _0\rightarrow \left[
\begin{array}{c}
\varphi _{\varepsilon ,\lambda } \\
-\varphi _{\varepsilon ,\lambda }
\end{array}
\right]  \label{17}
\end{equation}
where $\phi _{\varepsilon ,\lambda }$ and $\varphi _{\varepsilon
,\lambda }$ are $4$-spinors. $\lambda$ is set to $\pm 1/2$ and
refers to the two possible helicity states. From eq.10 the
solutions of $\Psi $ can then be easily derived

$\bullet E=E_p$ and $\varepsilon =+1$
\begin{equation}
u_\lambda ({\bf p)}=\frac 1{2\sqrt{E_p(E_p+m)}}\left(
\begin{array}{c}
\left( E_p+m\right) R\chi _\lambda \\
2\lambda pR\chi _\lambda \\
\left( E_p+m\right) R\chi _\lambda \\
2\lambda pR\chi _\lambda
\end{array}
\right)  \label{18}
\end{equation}

$\bullet E=E_p$ and $\varepsilon =-1$

\begin{equation}
v_\lambda ({\bf p})=\frac 1{2\sqrt{E_p(E_p+m)}}\left(
\begin{array}{c}
-2\lambda pRi\sigma _2\chi _\lambda \\
\left( E_p+m\right) Ri\sigma _2\chi _\lambda \\
-2\lambda pRi\sigma _2\chi _\lambda \\
\left( E_p+m\right) Ri\sigma _2\chi _\lambda
\end{array}
\right)  \label{19}
\end{equation}

$\bullet E=\widetilde{E}_p$ and $\varepsilon =+1$

\begin{equation}
\widetilde{u}_\lambda ({\bf p})=\frac 1{2\sqrt{\widetilde{E}_p(\widetilde{E}%
_p+m)}}\left(
\begin{array}{c}
(\widetilde{E}_p+m)R\chi _\lambda \\
(2\lambda p{\bf +}i2g)R\chi _\lambda \\
-(\widetilde{E}_p+m)R\chi _\lambda \\
-(2\lambda p{\bf +}i2g)R\chi _\lambda
\end{array}
\right)  \label{20}
\end{equation}

$\bullet E=\widetilde{E}_p$ and $\varepsilon =-1$%
\begin{equation}
\widetilde{v}_\lambda ({\bf p})=\frac 1{2\sqrt{\widetilde{E}_p(\widetilde{E}%
_p+m)}}\left(
\begin{array}{c}
(-2\lambda p{\bf +}i2g)Ri\sigma _2\chi _\lambda \\
(\widetilde{E}_p+m)Ri\sigma _2\chi _\lambda \\
-(-2\lambda p{\bf +}i2g)Ri\sigma _2\chi _\lambda \\
-(\widetilde{E}_p+m)Ri\sigma _2\chi _\lambda
\end{array}
\right)  \label{21}
\end{equation}
where $\chi _\lambda $ is such that $\chi _{1/2}=(1,0),$ and $\chi
_{-1/2}=(0,1)$. We have also
\begin{equation}
R=\exp \left[ -\frac i2\sigma _2\theta \right] \label{22}
\end{equation}

and
\begin{equation}
{\bf p}=\left( p\sin \theta ,0,p\cos \theta \right) \label{23}
\end{equation}

Note that the solutions given by eq.18 and 19 have been derived
assuming that the charge conjugation operator is given by
$C=i\Gamma ^2\Gamma ^0$ as a natural extension of its one sheeted
counterpart. Note that $E=E_p$ refers to symmetric states, whereas
$E=\widetilde{E}_p$ refers to antisymmetric states.

\section{Free field and fermion oscillations}

Using the 8-spinors solutions given by eq.16 and eq.18 we can now
try to build states corresponding to particles localized in a
specific sheet. For instance, a convenient state relative to a
particle localized initially in the $(+)$ sheet is given by

\begin{eqnarray}
\Psi (x) &=&\frac 1{\sqrt{2}\sqrt{V}}(N_{1/2}\left[ u_{1/2}({\bf p)}%
e^{-ip.x}+\widetilde{u}_{1/2}({\bf p)}e^{-i\widetilde{p}.x}\right]
\nonumber
\label{A13} \\
&&{\bf +}N_{-1/2}\left[ u_{-1/2}({\bf p)}e^{-ip.x}+\widetilde{u}_{-1/2}({\bf %
p)}e^{-i\widetilde{p}.x}\right] )  \label{24}
\end{eqnarray}
The polarization $P_e$ for such a particle is defined as
\begin{equation}
P_e=\frac{N_{1/2}^2{\bf -}N_{-1/2}^2}{N_{1/2}^2{\bf +}N_{-1/2}^2}=N_{1/2}^2%
{\bf -}N_{-1/2}^2  \label{25}
\end{equation}
such that $-1<P_e<1$ where $N_{1/2}^2+N_{-1/2}^2=1.$

To illustrate the basic predictions of the model, let us consider
the case of a fully polarized particle with a positive energy
\begin{equation}
\psi =\frac 1{\sqrt{2}}\left( u_\lambda ({\bf p)+}\widetilde{u}_\lambda (%
{\bf p})\right)  \label{26}
\end{equation}
It can be verified that in the limit of zero coupling, this states
reduces to
\begin{equation}
\lim_{g\rightarrow 0}\psi =\frac 1{\sqrt{2}}\frac
1{\sqrt{E_p(E_p+m)}}\left(
\begin{array}{c}
(E_p+m)R\chi _\lambda \\
2\lambda pR\chi _\lambda \\
0 \\
0
\end{array}
\right)  \label{27}
\end{equation}

such that there is no field contribution in the $(-)$ sheet. In
that case, the spinor takes the form of the usual solution of the
Dirac equation. At the first order of approximation, however, one
gets

\begin{equation}
\psi \sim \frac 1{\sqrt{2}}\frac 1{\sqrt{E_p(E_p+m)}}\left(
\begin{array}{c}
(E_p+m)R\chi _\lambda \\
(2\lambda p{\bf +}ig)R\chi _\lambda \\
0 \\
{\bf -}igR\chi _\lambda
\end{array}
\right)  \label{28}
\end{equation}

Now, one can see that the field component in the $(-)$ sheet is no
more strictly equals to zero. Such a result has several
interesting consequences for the particle. As an example, let us
consider the probability $P(t)$ for a particle of positive energy
to be localized in the $(+)$ sheet. We need to consider the first
four components of the spinor $\Psi $, i.e. the spinor $\psi _{+}$
and integrate $\left| \psi _{+}\right| ^2$ through the whole space
of the $(+)$ sheet. Then, one gets from eq.22

\begin{equation}
P(t)=\frac 12\left[ 1+A\cos \left[ (\widetilde{E}_p-E_p)t\right]
+B\sin \left[ (\widetilde{E}_p-E_p)t\right] \right] \label{29}
\end{equation}
where
\begin{equation}
A=\frac{(E_p+m)(\widetilde{E}_p+m)+p^2}{2\sqrt{E_p(E_p+m)}\sqrt{\widetilde{E}%
_p(\widetilde{E}_p+m)}}  \label{30}
\end{equation}
and
\begin{equation}
B=\frac{gpP_e}{\sqrt{E_p(E_p+m)}\sqrt{\widetilde{E}_p(\widetilde{E}_p+m)}}
\label{31}
\end{equation}

The form of $P(t)$ indicates that the particle oscillates between
the two sheets. Assuming $g\ll E_p$, the period of oscillations
$T_o$ can be expressed as

\begin{equation}
T_o=\frac \pi {g^2}E_p\left( 1+g\frac{pP_e}{2\pi E_p(E_p+m)}+{\cal O}%
[g]^3\right)  \label{32}
\end{equation}

It is worth noticing that particles of high energy undergo
oscillations of larger period than particles of low energy. This
is a very interesting result suggesting that contrarily to what
happens usually in branes models, the oscillations are strongly
suppressed for highly massive or energetic particles. Obviously
such oscillations would be observed from the perspective of a
brane observer as a violation of conservations law. But, in fact,
no violation occurs from a 5D point of view since the sum of the
energy on both sheets remain constant. The fact that such a
process has not been observed yet suggests a very weak coupling
constant g. In the limit where $g \rightarrow 0$, the two sheets
are completely decoupled and no oscillation occurs. We would like
to stress that $P(t)$ is not invariant through $P$ or $T$
transformations besides it is $PT$ invariant. Notice that the
results and considerations reported in this paragraph hold for a
negative energy particle as well.

For illustrative purpose, let us consider an electron such that
$p=0$. We are going to assume a half period of oscillations of the
order of the estimated proton lifetime , i.e. : $T_{o}/2\sim
10^{34}$ years. Then one gets $g\sim 2\times 10^{-19}$ m$^{-1}$.
This value corresponds to a separation distance between both
sheets of about $510$ l.y.. This is a particularly huge value in
comparison with the usual distances considered in branes theories.
Alternatively, assume that $\delta \sim 10^{-3}$ m and let us
consider an electron with a kinetic energy of $1$ keV. Then, the
half period of oscillation becomes $T_{o}/2 \sim 10^{-2}$ s which
corresponds to a travel distance of about $187$ km. If the kinetic
energy of the electron is subsequently decreased, i.e. $E \sim 25$
meV, the covered distance is still $1$ km. So, in both cases, the
disappearance of the particle into the other sheet is a phenomenon
that can not be easily observed. One may suggest to use a beam of
particles instead of individual particles to reveal the
oscillations. Nevertheless, interactions between particles in a
beam cannot be neglected anymore and strong suppression of the
oscillations will likely occur in that case. We will return to
this important problem later on in the paper. At last, these
oscillations for a free particle are typically a consequence of
the present ''finite difference`` approach. In the paper cited
before [18] where non-commutativity was used, these oscillations
do not appear at all.

\section{Incorporation of an electromagnetic field into the model}

In this section, the electromagnetic field will be introduced into
the two-sheeted Dirac equation. The choice of the electromagnetic
force, by contrast to electroweak or strong force, rests on the
choice of a simplified case of gauge group and just serves as an
illustration. One can easily convince oneself that the results
presented here for electromagnetism can be extended to the case of
other interactions like electroweak interaction or chromodynamics
for instance.

Logically, each sheet possesses its own current and charge density
distribution. It is also considered that photons can not travel
from one sheet to the other one. That means that each
electromagnetic field is confined within its own brane from which
it is native. In each brane, the electromagnetic field is then
described by the corresponding 4-vector potential $A_\mu ^{+}$ and
$A_\mu ^{-}$, each one being independent from the other one.

Classically, the electromagnetic field satisfies the gauge
condition

\begin{equation}
A_\mu ^{\prime }=A_\mu +\partial _\mu \Lambda  \label{33}
\end{equation}
in order for the electromagnetic field tensor
\begin{equation}
F^{\mu \nu }=\partial _\nu A_\mu -\partial _\mu A_\nu  \label{34}
\end{equation}
to be invariant through a gauge transformation. These conditions
can be easily extended to a five dimensional problem, even in the
case of a discretized extra-dimension. Usually, $\Lambda$ depends
of $x_\mu$ coordinates. What is going on then if $\Lambda$ also
depends of the extra-dimension? It will imply the existence of a
fifth potential vector components $A_5$. Keep safe actual
electromagnetic laws implies then that $A_5^{\pm }=0$, i.e. we
must impose that $\Lambda $ is independent of the fifth space
dimension. Indeed, from eq.31 one must verify that

\begin{equation}
A_5^{\prime \pm }=A_5^{\pm }+\left( \partial _5\Lambda \right)
^{\pm }=0  \label{35}
\end{equation}

and using the definition of the discrete derivative along the
extra dimension, this last equation corresponds to

\begin{equation}
\pm \frac 1\delta \left( \Lambda _{+}-\Lambda _{-}\right) =0
 \label{36}
\end{equation}
i.e. $\Lambda _{+}=\Lambda _{-}=\Lambda $. Considering the
two-branes wave function, the previous result means that we have a
gauge transformation $\exp (iq\Lambda )$ which must be applied to
both spacetime sheets simultaneously. This result could seem to be
in conflict with locality of the gauge transformation. In fact,
the gauge is global from the point of view of both branes
locations only. Of course, the locality is conserved according to
the $x_\mu$ coordinates and we should not be surprised about this
result which keeps unchanged the electromagnetic potential
properties. It allows to introduce two copies of the
electromagnetic field, each confined to a single brane, without
introducing a new electromagnetic potential component which would
be difficult to easily reconcile with actual observations. A
similar hypothesis was assumed in the NC approach of the
two-sheeted spacetime [18].

Now, the incorporation of the gauge field in eq.4 leads to

\begin{equation}
\left\{ i\Gamma ^\mu (\partial _\mu +iq{\bf A}_\mu )+ig\Gamma
^5-m\right\} \Psi =0   \label{37}
\end{equation}
where
\begin{equation}
{\bf A}_\mu =\left[
\begin{array}{cc}
A_\mu ^{+} & 0 \\
0 & A_\mu ^{-}
\end{array}
\right]   \label{38}
\end{equation}
corresponding to a Lagrangian density given by
\begin{equation}
{\cal L}=\overline{\Psi }\left\{ i\Gamma ^\mu \partial _\mu
+ig\Gamma ^5-m\right\} \Psi -q\overline{\Psi }\Gamma ^\mu \Psi
{\bf A}_\mu   \label{39}
\end{equation}
In the following, we will restrict ourselves to the case where the
kinetic energy is much smaller than the rest energy $m$. In that
case, we will show that it is possible to derive a Pauli like
equation valid for the two-sheeted spacetime.

\section{Extended Pauli equation}

To clarify the effect of the electromagnetic fields, let us
determine the non-relativistic limit of eq.35. Inspired from the
classical treatment, we are now looking for an equation satisfying

\begin{equation}
i\hbar \partial _t\chi ={\bf H}\chi  \label{40}
\end{equation}
with
\begin{equation}
\chi =\left[
\begin{array}{c}
\chi _{+} \\
\chi _{-}
\end{array}
\right]   \label{41}
\end{equation}
where $\chi _{+}$ and $\chi _{-}$ are two component spinors
related to the wave functions of the two sheets.

Eq.35 can be written as
\begin{equation}
i\partial _0\Psi =-i\Gamma ^0\Gamma ^\eta (\partial _\eta +iq{\bf
A}_\eta )\Psi -ig\Gamma ^0\Gamma ^5\Psi +m\Gamma ^0\Psi +q{\bf
A}_0\Psi  \label{42}
\end{equation}
When $m$ is large compared with the kinetic energy, the most rapid
time dependence arises from a factor $\exp (\pm imt)$. For a free
positive energy particle, the general solution is then given by
the product of $\exp (-imt)$ by one of the solutions (16) and
(18). For small kinetic and electromagnetic energies, therefore,
we look for solutions of the form
\begin{equation}
\Psi =\left[
\begin{array}{c}
\chi _{+} \\
\theta _{+} \\
\chi _{-} \\
\theta _{-}
\end{array}
\right] e^{-imt}  \label{43}
\end{equation}
where $\chi _{+}$, $\theta _{+}$, $\chi _{-}$, $\theta _{-}$ are
two-component spinors.

Then, the two-sheeted Dirac equation leads to the following system
of coupled differential equations

\begin{eqnarray}
i\partial _0\chi _{+} &=&-i\sigma _\eta \left( \partial _\eta
+iqA_\eta ^{+}\right) \theta _{+}+qA_0^{+}\chi _{+}-ig\left(
\theta _{+}-\theta _{-}\right)  \label{44}
\end{eqnarray}
\begin{equation}
i\partial _0\chi _{-}=-i\sigma _\eta \left( \partial _\eta
+iqA_\eta ^{-}\right) \theta _{-}+qA_0^{-}\chi _{-}+ig\left(
\theta _{+}-\theta _{-}\right)   \label{45}
\end{equation}
\begin{equation}
i\partial _0\theta _{+}=-i\sigma _\eta \left( \partial _\eta
+iqA_\eta ^{+}\right) \chi _{+}+qA_0^{+}\theta _{+}+ig\left( \chi
_{+}-\chi _{-}\right) -2m\theta _{+}   \label{46}
\end{equation}
\begin{eqnarray}
i\partial _0\theta _{-} &=&-i\sigma _\eta \left( \partial _\eta
+iqA_\eta ^{-}\right) \chi _{-}+qA_0^{-}\theta _{-}-ig\left( \chi
_{+}-\chi _{-}\right) -2m\theta _{-}   \label{47}
\end{eqnarray}

Since the mass $m$ is large in comparison with the kinetic energy
and coulomb terms, the components $\theta _{+}$ and $\theta _{-}$
are small and can be approximated by

\begin{equation}
\theta _{+}\approx -i\frac 1{2m}\sigma _\eta \left( \partial _\eta
+iqA_\eta ^{+}\right) \chi _{+}+i\frac g{2m}\left( \chi _{+}-\chi
_{-}\right)  \label{48}
\end{equation}
\begin{equation}
\theta _{-}\approx -i\frac 1{2m}\sigma _\eta \left( \partial _\eta
+iqA_\eta ^{-}\right) \chi _{-}-i\frac g{2m}\left( \chi _{+}-\chi
_{-}\right)  \label{49}
\end{equation}
By substituting eq.46 and eq.47 into eq.42 and eq.43, one finds

\begin{eqnarray}
i\partial _0\chi _{+} &=&-\frac 1{2m}\sigma _\eta \sigma _\nu \left( {\bf %
\nabla }-iq{\bf A}_{+}\right) ^\eta \left( {\bf \nabla }-iq{\bf A}%
_{+}\right) ^\nu \chi _{+}+q\Phi _{+}\chi _{+}  \nonumber  \\
&&+\frac{g^2}m\left( \chi _{+}-\chi _{-}\right) +iq\frac g{2m}{\bf
\sigma \cdot }\left\{ {\bf A}_{+}-{\bf A}_{-}\right\} \chi _{-}
\label{50}
\end{eqnarray}
\begin{eqnarray}
i\partial _0\chi _{-} &=&-\frac 1{2m}\sigma _\eta \sigma _\nu \left( {\bf %
\nabla }-iq{\bf A}_{-}\right) ^\eta \left( {\bf \nabla }-iq{\bf A}%
_{-}\right) ^\nu \chi _{-}+q\Phi _{-}\chi _{-}  \nonumber \\
&&-\frac{g^2}m\left( \chi _{+}-\chi _{-}\right) -iq\frac g{2m}{\bf
\sigma \cdot }\left\{ {\bf A}_{+}-{\bf A}_{-}\right\} \chi _{+}
\label{51}
\end{eqnarray}
where $\Phi _{\pm }=A^{0,\pm }$ and ${\bf A}_{\pm ,\eta }=A^{\eta
,\pm }$ are the usual electric potential and magnetic vector
potential. The Pauli matrices satisfying the identity
\begin{equation}
\sigma _i\sigma _j=\delta _{ij}+i\varepsilon _{ijk}\sigma _k
\label{52}
\end{equation}
where $\varepsilon _{ijk}$ is the Levi-Civita symbol, one gets :
\begin{equation}
\sigma _\eta \sigma _\nu \left( {\bf \nabla }-iq{\bf A}_{\pm
}\right) ^\eta
\left( {\bf \nabla }-iq{\bf A}_{\pm }\right) ^\nu =\left( {\bf \nabla }-iq%
{\bf A}_{\pm }\right) ^2+q{\bf \sigma \cdot B}_{\pm } \label{53}
\end{equation}

The Hamiltonian of eq.38 is then given by a sum of several
different terms
\begin{equation}
{\bf H}={\bf H}_k{\bf +H}_m{\bf +H}_p{\bf +H}_c{\bf +H}_{cm}
\label{54}
\end{equation}
which are (in natural units) :
\begin{eqnarray}
{\bf H}_k=-\frac{\hbar ^2}{2m}\left[
\begin{array}{cc}
\left( {\bf \nabla }-i\frac q\hbar {\bf A}_{+}\right) ^2 & 0 \\
0 & \left( {\bf \nabla }-i\frac q\hbar {\bf A}_{-}\right) ^2
\end{array}
\right]  \label{55}
\end{eqnarray}
\begin{eqnarray}
{\bf H}_m=-g_s\mu \frac \hbar 2\left[
\begin{array}{cc}
{\bf \sigma \cdot B}_{+} & 0 \\
0 & {\bf \sigma \cdot B}_{-}
\end{array}
\right]   \label{56}
\end{eqnarray}
\begin{eqnarray}
{\bf H}_p=\left[
\begin{array}{cc}
q\Phi _{+}+V_{+} & 0 \\
0 & q\Phi _{-}+V_{-}
\end{array}
\right]   \label{57}
\end{eqnarray}
\begin{equation}
{\bf H}_c=\frac{g^2\hbar ^2}m\left[
\begin{array}{cc}
1 & -1 \\
-1 & 1
\end{array}
\right]  \label{58}
\end{equation}
\begin{eqnarray}
{\bf H}_{cm}=igg_c\mu \frac \hbar 2\left[
\begin{array}{cc}
0 & {\bf \sigma \cdot }\left\{ {\bf A}_{+}-{\bf A}_{-}\right\} \\
-{\bf \sigma \cdot }\left\{ {\bf A}_{+}-{\bf A}_{-}\right\} & 0
\end{array}
\right]   \label{59}
\end{eqnarray}
with $\mu =q/2m$ the Bohr magneton and $g_s=g_c=2$.

The first three terms of the Hamiltonian (52), i.e. ${\bf H}_k$,
${\bf H}_m$ and $ {\bf H}_p$, remind the usual terms of the
classical Pauli equation in presence of an electromagnetic field :
${\bf H}_k$ relates to the kinetic part of the hamiltonian taking
account the vector potential. ${\bf H}_m$ is the coupling term
between the magnetic field and the magnetic moment of the particle
$g_s\mu $ with $g_s$ the gyromagnetic factor and $ {\bf H}_p$ is
the coulomb term. In addition to these terms, the hamiltonian
contains a coupling term $ {\bf H}_c$ linking the two sheets
together and which was previously responsible of the oscillations
of the free particle. Note that except this last term and some
minor differences regarding the notations (as the use of natural
units and the explicit introduction of the magneton) the extended
Pauli equation derived in this paper is actually the same that the
one obtained using a NC formalism [18]. Hence, we see confirm that
a classical treatment using finite differences in discrete space
permits to
reproduce quite easily the NC results.\\
So, ${\bf H}_{cm}$ introduces a pure electromagnetic coupling term
involving the magnetic vector potential and something like a
magnetic moment given by $g_c\mu $, where $g_c$ is analogous to
the gyromagnetic factor. Of course, we still have $g_s=g_c=2$. At
that point, it seems important to remind that $g_s$ is not
strictly equals to $2$ due to vacuum effects accurately predicted
by QED. So, there is no certainty that $g_c\mu $ closely
corresponds to the magnetic moment $g_s\mu $. In doubt, we now
refer $g_c\mu $ as being the isomagnetic moment and $g_c$ the
isogyromagnetic factor by analogy with $g_s\mu$ the magnetic
moment and $g_s$ the gyromagnetic factor. For the proton and the
neutron, one usually uses the nuclear magneton (which is defined
by the mass and the charge of the proton instead of the values of
the electron). Then, the $g_s$ factors derived using the classical
Dirac equation are 2 and 0 respectively. There is a large
discrepancy between the predicted values and the experimental ones
which are 5.58 and -3.82 respectively. It is well known that the
difference arises from the fact that the electron is a fundamental
particle whereas proton and neutron which are composed by quarks,
are not. As a consequence, the Dirac equation which applies
normally only to point like particles cannot be applied directly
to the proton and the neutron. Practically, the use of the Dirac
equation to describe these particles requires the use of the
experimental values of the gyromagnetic factors $g_s$ thus
defining the anomalous magnetic moment. In the same spirit, one
can expect that the $g_c$ factors are not exactly 2 and 0 for the
proton and the neutron and we may assume something like an
anomalous isomagnetic moment as well. If this assumption holds,
then it would mean that the coupling ${\bf H}_{cm}$ between the
two sheets occurs also for the neutron despite its zero electrical
charge.

It could be objected that neutron oscillations are suspicious
since it neglects the internal structure of the particle made of
quarks always interacting \textit{via} gluons exchange. But, as
for the electromagnetic force, we assumed that the strong force
exists as two copies of the gluons fields each one confined in its
own brane. In our model, only the quarks, which are fermions,
would be able to oscillate from one brane to the other, contrary
to the gluons. So, what about the neutron? The quarks form a
strongly bounded and entangled system and they must oscillate
together. As the neutron oscillates, at each time, it is
delocalized on each brane at the same time. Of course, it is the
same thing for the related quarks bag. A related assumption is
that the cohesion of a delocalized bag in one brane is then
ensured by the gluons field associate with this brane. In this
way, as the neutron is transferred, the quarks are transferred,
and the role of the gluons field of the first brane is
progressively substituted by the gluons field of the second brane.
Note that similar considerations can also be applied to the NC
approach of the problem [18].

In the following we shall admit these assumptions as true and
consider two cases of coupling with the aim to highlight some
basic features of our model.

\section{Neutron in a constant scalar potential}

Let us consider the case of a single neutron embedded in a region
of constant potential in the $(+)$ sheet.  In the rest frame of
the particle, the Hamiltonian ${\bf H}$ reduces to :

\begin{equation}
{\bf H}=\left[
\begin{array}{cc}
V_0 & 0 \\
0 & 0
\end{array}
\right] +\frac{g^2\hbar ^2}m\left[
\begin{array}{cc}
1 & -1 \\
-1 & 1
\end{array}
\right]  \label{60}
\end{equation}
If one assumes that the particle is originally located in the
$(+)$ sheet, a simple calculus based on eq.38 gives the
probability to find the particle in the second sheet. One gets :
\begin{equation}
{\cal P}_{inv}=\frac 1{1+\gamma ^2}\sin ^2\left( \hbar ^{-1}\beta \sqrt{%
1+\gamma ^2}t\right) \label{61}
\end{equation}
where $\beta =g^2\hbar ^2/m$ and $\gamma =V_0/(2\beta )$.

When $\gamma =0$, i.e. without any potential, the particle
oscillates freely between the two sheets with a period $T=\pi
m/(g^2\hbar )$ similar to the one found previously (cf. eq.30).
But, when the potential is switched on, the period of oscillations
changes. In fact, the greater the potential is, the lower the
period is. Moreover, the amplitude of the oscillations drops very
quickly with the potential. Such a result shows that the field
induces a confinement of the particle in its sheet. Perhaps, such
a mechanism could be responsible for the matter stability even for
a large coupling constant $g$. Let us assume for instance that
$g=10^3$, assuming a neutron with a kinetic energy of about $25$
meV, the period is now of the order $T\sim 50$ s which is a too
small value to be compatible with the matter stability.
Nevertheless, the corresponding value of beta is $\beta \sim
4\times 10^{-17}$ eV which is quite small. By contrast, the lowest
temperatures obtained in laboratory are around $T_{emp}\sim
50\times 10^{-9}$ K and correspond to energy of the order
$k_{B}T_{emp}\sim4\times 10^{-12}$ eV. Even a value of $V_0$
corresponding to such a low energy is still associated with a
$\gamma \sim 5\times 10^4$, i.e. a maximum probability amplitude
of $4\times 10^{-10}$. It is thus clear that even if the coupling
constant is large, the particles are quickly and strongly confined
in their own sheet. Obviously, the environmental effects inhibits
dramatically the particle oscillations between the two sheets. So
we see that even if the present approach predicts free
oscillations by contrast to the NC approach where particles are
perfectly stable in their own sheet, the confinement discussed
here leads to a similar physical result: in a usual physical
environment, particles do not oscillate at all, they are glued in
the brane.

\section{Case of a neutron embedded in a region of constant magnetic vector potential}

In our previous paper [18], it was shown that the particle
oscillations could be enhanced in some appropriate situations
involving intense magnetic potentials. Let us now demonstrate that
this result can also be obtained with the present approach. As
previously, we are considering the simplified case of a neutron,
assuming that $g_c$ is not strictly equals to zero as suggested
previously. We make the complementary assumption $g_c\sim g_s$.
Assume that there is a constant magnetic vector potential in the
$(+)$ sheet in the region where the particle is initially located.
Such a potential could be experimentally realized for instance, by
considering a uniform current map along a hollow cylinder. If the
current intensity is $I$ then the magnetic vector potential ${\bf
A}$ appearing inside the hollow part of the cylinder has a module
of the order $A\sim \mu _0I$.

In the Hamiltonian given by eq.52, the eigenvalues of ${\bf H}_c$
are $2g^2\hbar ^2/m$ and $0$, each one being doubly degenerated.
The eigenvalues of ${\bf H}_{cm}$ are $\pm (1/2)gg_c\mu \hbar
\left| {\bf A}\right| $ and are also doubly degenerated. The
typical order of magnitude for the energies related to ${\bf H}_c$
and ${\bf H}_{cm}$ are $E_c$ and $E_{cm}$ respectively such that
the ratio between both contributions is approximately
$E_{cm}/E_c\sim \left| q\right| \left| {\bf A} \right| /(g\hbar
)$. If we set $\left| {\bf A}\right| \gg g\hbar /\left| q\right| $
then we get $E_{cm}\gg E_c$ such that ${\bf H}_c$ can now be
treated as a weak perturbation of ${\bf H}_{cm}$. For a small
coupling constant $g$, this condition is not very restrictive and
we may assume that it could be achieved practically. Thus to
simplify further the calculations, we are neglecting ${\bf H}_c$.
In such conditions, the hamiltonian reduces to (in the rest frame
of the neutron and with ${\bf A=}A{\bf e}$) :

\begin{equation}
{\bf H}=\left[
\begin{array}{cc}
V_0 & 0 \\
0 & 0
\end{array}
\right] {\bf +}i\hbar \Omega \left[
\begin{array}{cc}
0 & {\bf \sigma \cdot e} \\
-{\bf \sigma \cdot e} & 0
\end{array}
\right]  \label{62}
\end{equation}
where $\Omega =(1/2)gg_s\mu A$ with
\begin{equation}
{\bf e=}\left[
\begin{array}{c}
\sin \theta \cos \varphi  \\
\sin \theta \sin \varphi  \\
\cos \theta
\end{array}
\right]   \label{63}
\end{equation}
Provided that the neutron is originally located in the sheet $(+)$
and considering that $(\theta ,\varphi )$ gives the relative
direction between ${\bf A}$ and the spin, it is straightforward to
derive the transfer probability, i.e.
\begin{equation}
{\cal P}_{inv}=\frac 1{1+\widetilde{\gamma }^2}\sin ^2\left( \Omega \sqrt{1+%
\widetilde{\gamma }^2}t\right)  \label{64}
\end{equation}
with $\widetilde{\gamma }=V_0/(2\Omega \hbar ).$

Eq.62 exhibits similar properties to the one derived in eq.59.
Nevertheless, the magnetic vector potential plays now the role of
a coupling constant which supplements $g$. Thus, the probability
of transfer is now directly related to the value of $A$ which can
be controlled experimentally. Note that eq.62 is consistent with
the one derived using a NC formalism (see eq.51 in the
aforementioned paper). Hence, from the point of view of artificial
oscillations, both approaches are in perfect agreement. It is
useful to define a critical value for $A$ given by $A_c=V_0/(\hbar
gg_s\mu )$ insuring that the maximum probability amplitude would
be equal to $1/2$ at least. This value is defined in accordance
with the estimated value of $V_0$ which can be seen, in this
idealized case as an indicator of the environmental effects. Let
us assume for instance that $g=10^3$. In a typical cooled
($T_{emp}\sim 1$ K) and insulated environment, one can expect for
$V_0\sim k_{B}T_{emp}\sim 86$ $\mu$eV leading to $A_c\sim 0.72$
T.m. The related current intensity required to satisfy these
conditions is $I\sim 0.6\times10^{6}$ A. At room temperature, one
can consider instead $V_0\sim 25$ meV corresponding to a value of
$A_c$ around $207.5$ T.m. The related current intensity to produce
oscillations now becomes $I\sim 0.2\times 10^9$ A. At first
glance, one may think that the most simple way to produce
artificial oscillations would be to use a cooled and insulated
device with a neutron beam flowing inside. However, one must keep
in mind that the coldest neutron beams still correspond to
energies about $25$ meV. In such circumstances, the interactions
between neutrons in the beam could likely inhibit the oscillations
and force the particles to stay in their own spacetime sheet. A
possible solution could be to use a weakly intense source of cold
neutrons in order to prevent their interactions. As an example, a
typical experimental device should be constituted by a one-by-one
ultra cold neutron source ($25$ meV) and a conducting cylinder
with a current intensity $I\sim 0.6\times10^{6}$ A. Of course, the
experimental device should be placed in a cooled environment at a
temperature of about $1$\ K. In such conditions, the neutrons
would exhibit a typical velocity of about $2187$ m.s$^{-1}$ and
the oscillations should take place with a half period of about
$17$ ps only. That means that neutrons could disappear in the
other spacetime sheet (with a maximum probability amplitude of
$1/2$) after covering a distance of about $37$ nm.

\section{Conclusion}

In this paper, Dirac and Pauli like equations valid for a
two-sheeted spacetime have been derived and studied. Our model can
be adequately used to mimic two-branes universe at low energies
and thus could be relevant for the study of branes phenomenology.
The results of this work are almost identical to those obtained in
a previous paper [18] where a similar idea of quantum two-sheeted
spacetime was developed using NCG. The mathematics used in this
paper present the advantage of being more appealing since they are
more simple. Nevertheless, several minor differences between both
approaches have been noticed. Contrarily to the non-commutative
approach, it is shown that a transfer of matter can occur from one
brane to the other one even for free particles. This is a
specificity of the ''finite difference`` analysis of the present
paper. Concerning the effect of electromagnetic field, no
significant difference between both approaches have been noted. We
have demonstrated the existence of an oscillatory behavior of the
particles in the presence of magnetic potentials but a freezing of
the oscillations in presence of scalar potentials. In this paper,
a possible experimental set-up, relevant to force particles like
neutrons to oscillate, has been proposed. It is suggested that
some specific configurations of electromagnetic fields, accessible
with our present technology, could be adequately used to achieve
this goal.


\begin{thebibliography}{19}

\bibitem{1}Th. Kaluza, ``Zum Unit\"{a}tsproblem der Physik'', Sitz. Preuss.
Akad. Wiss. Phys. Math. Kl. (1921) 966

\bibitem{2}O. Klein, ``Quantentheorie und funfdimensionaler
Relativit\"{a}tstheorie'', Z. Phys. 37 (1926) 895

\bibitem{3}R. Bryan, ``Are the Dirac particles of the Standard Model
dynamically confined states in a higher-dimensional flat space?'',
Can. J. Phys. 77 (1999) 197

\bibitem{4}L. Randall, R. Sundrum, ``Large mass hierarchy from a small extra
dimension'', Phys. Rev. Lett. 83, 3370 (1999)

\bibitem{5}L. Randall, R. Sundrum, ``An alternative to compactification'', Phys.
Rev. Lett. 83, 4690 (1999)

\bibitem{6}R. Gregory, V.A. Rubakov, S.M. Sibiryakov, ``Brane worlds: the
gravity of escaping matter'', Class. Quant. Grav. 17, 4437 (2000)

\bibitem{7}S.L. Dubovsky, V.A. Rubakov, P.G. Tinyakov, ``Brane world:
disappearing massive matter'', Phys. Rev. D 62, 105011 (2000)

\bibitem{8}A. Barvinsky, A. Kamenshchik, C. Kiefer, A. Rathke, ``Graviton
oscillations in the two-brane world'', Phys. Lett. B571, (2003)
229-234

\bibitem{9}A. Kokado, G. Kouisi, T. Saito, K. Uehara, ``Brans-Dicke theory on
$M4 \times Z2$ geometry'', Prog. Theor. Phys. 96 (1996) 1291-1300

\bibitem{10}N. Arkani-Hamed, M.D. Schwartz, ``Discrete Gravitational dimensions'',
Phys. Rev. D 69, 104001 (2004)

\bibitem{11}C. Deffayet, J. Mourad, ``Solutions of multigravity theories and
discretized brane worlds'', Class. Quant. Grav. 21 (2004)
1833-1848

\bibitem{12}C. Deffayet, J. Mourad, ``Multigravity from a discrete
extradimension'', Phys. Lett. B589 (2004) 48-58

\bibitem{13}N.A.Viet, K.C.Wali, ``Non-commutative geometry and a
Discretized Version of Kaluza-Klein theory with a finite field
content'', Int. J. Mod. Phys. A11 (1996) 533

\bibitem{14}A. Connes, Non-Commutative Geometry, [Academic Press, 1994].

\bibitem{15}J.M.Gracia-Bondia, B.Iochum {\&} T.Schucker, ``The standard
model in noncommutative geometry and fermion doubling'', Phys.
Lett. B416 (1998) 123-128

\bibitem{16}F.Lizzi, G.Mangano, G.Miele {\&} G.Sparano, ``Mirror Fermions
in Noncommutative Geometry'', Mod. Phys. Lett. A13 (1998) 231-23

\bibitem{17}N.A. Viet, K.C. Wali, ``Chiral spinors and gauge fields in
noncommutative curved space-time'', Phys. Rev. D 67, 124029 (2003)

\bibitem{18}F. Petit, M. Sarrazin, ``Quantum dynamics of massive particles in a
non-commutative two-sheeted space-time '', Phys. Lett. B612 (2005)
105-114 (hep-th/0409084)


\end{thebibliography}
\end{document}